\pdfoutput=1
\documentclass{article}

\usepackage[preprint]{neurips_2019}

\usepackage[utf8]{inputenc} 
\usepackage[T1]{fontenc} 
\usepackage{hyperref}       
\usepackage{url}            
\usepackage{booktabs}      
\usepackage{amsfonts}      
\usepackage{nicefrac}      
\usepackage{microtype}    
\usepackage{algorithmic}
\usepackage[ruled,vlined]{algorithm2e}
\usepackage{amsmath}
\usepackage{graphicx}
\usepackage[font=footnotesize, labelfont=bf]{caption}

\usepackage{hyperref}
\hypersetup{colorlinks=true,linkcolor=blue,urlcolor=blue}

\usepackage{tikz}

\usepackage{amssymb}
\usepackage{pgfplots}
\usepackage{pgfmath}
\usepgfplotslibrary{patchplots}
\usetikzlibrary{patterns, positioning, arrows}

\pgfmathsetmacro\sprayRadius{.75pt}
\pgfmathsetmacro\sprayPeriod{.8cm}

\pgfdeclarepatternformonly{spray}{\pgfpoint{-\sprayRadius}{-\sprayRadius}}{\pgfpoint{1cm + \sprayRadius}{1cm + \sprayRadius}}{\pgfpoint{\sprayPeriod}{\sprayPeriod}}{
    \foreach \x/\y in {2/53,6/52,11/48,23/49,20/47,32/46,41/47,47/51,56/52,46/44,4/43,16/42,33/41,41/37,49/35,55/31,37/35,44/30,28/37,24/36,17/37,7/38,0/31,8/29,18/31,28/30,37/28,30/27,46/24,51/21,24/23,12/24,4/21,18/19,12/16,31/21,38/18,26/16,46/16,56/12,52/10,45/8,51/4,37/12,35/7,24/9,14/9,2/12,8/6,15/4,27/0,34/1,40/1} {
    \pgfpathcircle{\pgfpoint{(\x) / 10 * \sprayPeriod}{\sprayPeriod - (\y) / 10 * \sprayPeriod}}{\sprayRadius}
   }
\pgfusepath{fill}
}
\newcommand*{\affaddr}[1]{#1} 
\newcommand*{\affmark}[1][*]{\textsuperscript{#1}}
\newcommand*{\email}[1]{\texttt{#1}}
\title{HHHFL: Hierarchical Heterogeneous Horizontal Federated Learning for Electroencephalography}

\author{
Dashan Gao\affmark[1] \thanks{Equal Contribution} \hspace{0.1em}  , Ce Ju\affmark[2] \footnotemark[1]  \hspace{0.1em} ,  \\ 
\textbf{Xiguang Wei\affmark[2], Yang Liu\affmark[2], Tianjian Chen\affmark[2] and Qiang Yang\affmark[1]}\\
\affaddr{\affmark[1]Hong Kong University of Science and Technology }\\
\affaddr{\affmark[2]AI Lab, WeBank Co., Ltd.}\\
\email{dgaoaa@connect.ust.hk}\\
\email{\{ceju, xiguangwei, yangliu, tobychen\}@webank.com}\\
\email{qyang@cse.ust.hk}
}

\begin{document}

\setcitestyle{numbers,square}
\maketitle

\begin{abstract}
Electroencephalography (EEG) classification techniques have been widely studied for human behavior and emotion recognition tasks. But it is still a challenging issue since the data may vary from subject to subject, may change over time for the same subject, and maybe heterogeneous. Recent years, increasing privacy-preserving demands poses new challenges to this task. The data heterogeneity, as well as the privacy constraint of the EEG data, is not concerned in previous studies. To fill this gap, in this paper, we propose a heterogeneous federated learning approach to train machine learning models over heterogeneous EEG data, while preserving the data privacy of each party. To verify the effectiveness of our approach, we conduct experiments on a real-world EEG dataset, consisting of heterogeneous data collected from diverse devices. Our approach achieves consistent performance improvement on every task. 
\end{abstract}

\section{Introduction}
Electroencephalography (EEG) is an electrophysiological monitoring method to record electrical activity of the brain \cite{chong2013parameter}. It is typically noninvasive, with the electrodes placed along the scalp. Studies have shown that EEG signals can effectively reflect a person's fatigue, panic, alertness, behavioral intentions, epilepsy, and other information \cite{subha2010eeg}. 
Therefore, EEG signals have a wide range of applications. For example, in education, EEG can be used to help inspect and improve the concentration of students \cite{angelakis2007eeg} and its application in neural-feedback is investigated as well to help students learn to control and change their brain activity \cite{Li:2009:TAL:1631111.1631118}. 

Many EEG studies utilize model-based approaches that a large amount of data is required to build an accurate and robust model. In these approaches, EEG data of different people are collected to a central server for further processing.
However, as EEG signals reflect brain activities in numerous aspects, the potential abuse of EEG data may lead to sever privacy violation. Such ethical and privacy concerns has attracted public attention \cite{yu2018building}. In the European Union, the General Data Protection Regulation (GDPR) \cite{regulation2016regulation} specified many terms for protecting user privacy and prohibit organizations from exchanging data without explicit user approval. Under the increasingly stringent data security and privacy protection legislation, it is hard to collect abundant data for any single device and it is significant to conduct joint EEG signal analysis while protecting user privacy. 

Federated learning (FL) \cite{DBLP:journals/corr/McMahanMRA16}
is an emerging but powerful technique to solve this kind of problem that it can jointly train a machine learning model using data on different clients while ensuring their data privacy. 
Existing federated learning mainly focus on homogeneous dataset, where the different parties share the same feature space. 

For EEG signal collection, even the same equipment manufacturer may develop EEG signal acquisition equipment with varying electrode number, position and sampling rate let alone different equipment vendors. Such device diversity further exacerbates the scarcity of training data and leads to numerous distributed heterogeneous datasets in EEG classification. 
Heterogeneous domain adaptation linking different feature spaces based on labels has been studied \cite{wang2011heterogeneous11}. However, existing solutions assume multiple source domains with abundant labeled instances and one target domain with limited labeled instances as well as unlabeled instances. When applied to our problem setting, existing heterogeneous domain adaptation approaches will face significant accuracy drop. The reason is that in our setting, there are numerous clients each with limited labeled data, any of them can hardly act as a source domain. Moreover, heterogeneous domain adaptation should be adapted multiple times by taking each party as the target domain every time.

\begin{figure}[h]
\begin{minipage}[b]{0.55\linewidth}
To address this challenge, we propose a novel hierarchical heterogeneous horizontal federated learning (HHHFL) approach to leverage the heterogeneous EEG signals collected from diverse devices. Each party plays a role of both source domain and target domain, thus is equal. So that the problems faced by existing heterogeneous domain adaptation approaches are eliminated. Knowledge is accumulated among all the clients and transferred to each other. Clients of the same type of device collaboratively train a feature mapping model to map the heterogeneous data into one common embedding space via horizontal FL. Meanwhile, all clients build an EEG signal classifier to conduct EEG signal classification over the homogeneous features in the common embedding space. We conduct extensive experiments over a EEG dataset \emph{MindBigData}, where data of the same task is collected from diverse devices. Experimental results demonstrate that our proposed approach outperforms the model built locally without FL significantly.
\end{minipage}
\begin{minipage}[b]{0.5\linewidth}
\centering
	\includegraphics[height = 0.3\textheight, width=0.75\textwidth]{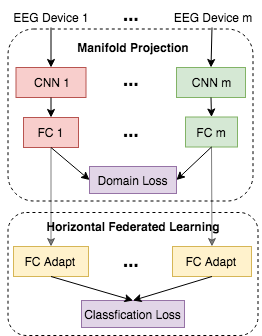}
	\caption{Hierarchical Architecture}
	\label{EEG}
\end{minipage}
\end{figure}

\section{Related Work}
EEG data classification has been extensively studied using various machine learning approaches \cite{hwang2013eeg,cnnSchirr17,bird2019deep,6335751}. Among which, Schirrmeister et al. \cite{cnnSchirr17} utilized convolutional neural networks (CNN) to distinguish pathological from normal EEG recordings and reaches state-of-the-art performance. Brid et al. \cite{bird2019deep} explored Multi-layer Perceptron (MLP) and Long Short-Term Memory (LSTM) augmented with adaptive boosting. 
Shenkai and Yaochu \cite{6335751} noticed the heterogeneous problem and proposed a heterogeneous classifier which consists different base classifiers as well as different feature extraction approaches for accurate EEG signal classification.
Although data collected from different devices for the same task has been studied for EEG signal classification \cite{roesler2014comparison}, as far as we known, there is no attempt to build a EEG classifier over heterogeneous EEG data collected from different devices, let along further taking the privacy issue into consideration. 

FL is a potential technique to solve the problem, it is first proposed by Google in 2016 \cite{DBLP:journals/corr/McMahanMRA16,DBLP:journals/corr/KonecnyMRR16} and it can train machine learning models under privacy constraint using decentralized data residing on end devices. A possible approach to bridge heterogeneous feature spaces in FL is discussed in \cite{liu2018secure}, it proposed a FTL approach which leverages instance co-occurrence in different parties and builds multiple correlated models instead of a single model. Also, federated multi-task learning is studied in \cite{NIPS2017_7029}, which deals with each task seperately. Symmetric transformation approaches \cite{wang2011heterogeneous11,wang2018heterogeneous18,duan2012learning} project source and target feature spaces into a domain-invariant feature subspace to associate cross-domain data, while most of them aim to boost the performance in one target domain.

\section{Problem Statement}
Each EEG device captures signals from multiple sensors located in different area of brain. The sensor readings are high-dimensional data distributed on some embedding submanifold $\mathcal{M} \hookrightarrow  \mathbb{R}^{d}$, where dimension $d$ is a large number.  

Assume we have $m$ cluster of sensor readings on $m$ embedding submanifold $\mathcal{M}_i  \hookrightarrow  \mathbb{R}^{d_i}$, $i = 1, \cdots, m$. The sensor readings are private and only accessed by users. We aim to find an privacy-preserving classifier to detect if one can recognize the digits in the image from electroencephalogram signals. 
\section{Methodology}
In this section, we propose our approach to find the privacy-preserving classifier. We take two strategies to prevent the high dimensionality and data privacy issues of sensor readings on EEG devices collected from multiple data source in Section \ref{projection} and Section \ref{Federated} respectively. The hierarchical architecture of our approach refers to Figure \ref{EEG} and Figure \ref{app} in Appendix.

\subsection{Manifold Projection}\label{projection}
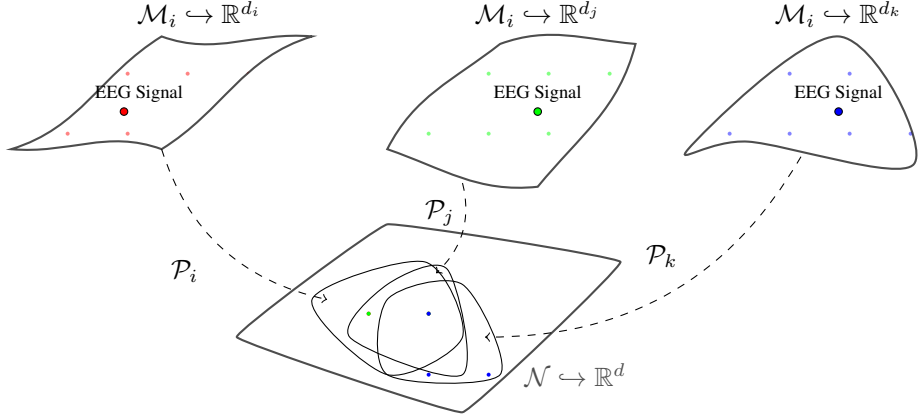
\begin{figure}
\centering
\begin{tikzpicture}
   
    \begin{scope}[out=-5, in=160, relative]
    \draw [thick] (2, 1.5) to (0, 0) to (2, 0) to (4, 1.5) to cycle[smooth cycle, tension=0.5, fill=white, pattern color=red, pattern=spray, opacity=0.7]; 
    \end{scope}
     \draw [fill=red](1.5, 0.5) circle (.05); 
     \fill[black, font=\scriptsize](1.7, 0.5) node [above] {EEG Signal};
     \fill[black](2.5, 1.5)  node [above] {$\mathcal{M}_i \hookrightarrow  \mathbb{R}^{d_i}$};

     \begin{scope}[out=-5, in=200, relative]
     \draw [thick] (6.5, 1.4) to (5, 0) to (7, -0.5) to (8.3, 1.4) to cycle[smooth cycle, tension=0.5, fill=white, pattern color=green, pattern=spray, opacity=0.7]; 
     \end{scope}
     \draw [fill=green](7, 0.5) circle (.05); 
     \fill[black, font=\scriptsize](7, 0.5) node [above] {EEG Signal};
     \fill[black](7, 1.5)  node [above] {$\mathcal{M}_i \hookrightarrow  \mathbb{R}^{d_j}$};
     
     \draw [thick] [smooth cycle, tension=0.6, fill=white, pattern color=blue, pattern=spray, opacity=0.7] plot coordinates{(11, 1.5) (9, 0) (9.8, 0.1) (12, -0.2)};
     \draw [fill=blue](11, 0.5) circle (.05); 
     \fill[black, font=\scriptsize](11, 0.5) node [above] {EEG Signal};
      \fill[black](11, 1.5)  node [above] {$\mathcal{M}_i \hookrightarrow  \mathbb{R}^{d_k}$};

    \draw [thick] [smooth cycle, tension=0.1, fill=white, pattern color=white, pattern=north west lines, opacity=0.7] plot coordinates{(8.1, -1.5) (5, -1) (3, -2.5) (6, -3.5)} node at (7.5, -3) {$\mathcal{N} \hookrightarrow  \mathbb{R}^{d}$};
    \draw[smooth cycle, pattern color=red, pattern=spray] 
        plot coordinates { (6, -2.5) (5.5, -1.5) (4, -2) (5, -3) } 
        node [label={[label distance=-0.3cm, xshift=-2cm, fill=white] }] {};
        
    \draw[smooth cycle, pattern color=green, pattern=spray] 
        plot coordinates { (6, -3) (5.8, -1.6) (5, -1.8) (4.5, -2.5) } 
        node [label={[label distance=-0.3cm, xshift=-2cm, fill=white] }] {};
        
     \draw[smooth cycle, pattern color=blue, pattern=spray] 
        plot coordinates { (6.5, -3) (6, -1.8) (5, -2) (5, -3) } 
        node [label={[label distance=-0.3cm, xshift=-2cm, fill=white] }] {};

    \path[dashed]  [->] (2, 0) edge [bend right] node[midway, xshift=-5mm, yshift=-3mm] {$\mathcal{P}_i$} (4.2, -2);
    \path[dashed] [->] (6, -0.45) edge [bend left] node[midway, xshift=-3mm, yshift=2.5mm] {$\mathcal{P}_j$} (5.65, -1.65);
    \path[dashed]  [->] (10.5, -0.1) edge [bend left] node[midway, xshift=-1mm, yshift=5mm] {$\mathcal{P}_k$} (6.3, -2.5);

\end{tikzpicture}
\caption{Manifold Projection}
\label{manifold}
\end{figure}

For each cluster $i$ of sensor readings $\{x_1, \cdots, x_{N_i}\}$ on $\mathcal{M}_i \hookrightarrow  \mathbb{R}^{d_i}$, we build its own projection $\mathcal{P}_i$ to map embedding submanifold $\mathcal{M}_i $ to the common embedding space $\mathcal{N} \hookrightarrow  \mathbb{R}^{d}$. Upon the establishment of manifold projections, we require the raw EEG data fall closely onto the common embedding space $\mathcal{N}$. We approximate these projections $\mathcal{P}_i$  with neural networks approach. The illustration of manifold projection refers to Figure \ref{manifold}.

\subsection{Loss}
The loss in our architecture is designed for minimizing both the classification loss and the domain loss as follows, 
\begin{itemize}
\item Classification Loss: $\mathcal{L}_C(X_{EEG}, Y)$ is the typical classification loss (i.e. cross entropy) over the whole dataset, $X_{EEG}$, and the ground truth labels, $Y$.

\item Domain Loss: We apply maximum mean discrepancy (MMD) \cite{gretton2007kernel} to measure the distances between probability distributions of projected EEG data on Manifold $\mathcal{N}$. Suppose each projected data $\mathcal{P}_i \big(\{x_1, \cdots, x_{N_i}\} \big) \sim \mathcal{Q}_i$, where $\mathcal{Q}_i$ is the probability distribution over embedding space $\mathcal{N}$. For a feature map $\psi: \mathcal{N} \longrightarrow \mathcal{H}$, where $\mathcal{H}$ is a reproducing kernel Hilbert space. The MMD in our case is defined as follows, 
\[
\text{MMD}(\mathcal{Q}_i, \mathcal{Q}_j) := || \mathbb{E}_{ \mathcal{P}_i(X_i) \sim\mathcal{Q}_i} \psi( \mathcal{P}_i(X_i) ) -  \mathbb{E}_{ \mathcal{P}_j(X_j)\sim\mathcal{Q}_j} \psi( \mathcal{P}_j(X_j) ) ||_{\mathcal{H}},
\]
where $X_i$ is the local cluster $i$ of sensor readings $\{x_1, \cdots, x_{N_i}\}$.
\end{itemize}

The overall loss is the sum of two kinds upon the two criteria as follows, 
\[
\mathcal{L} := \mathcal{L}_C(X_{EEG}, Y) + \sum_{1 \leq i < j \leq m} \lambda_{i, j} \cdot \text{MMD}^2(\mathcal{Q}_i, \mathcal{Q}_j).
\]


\subsection{Federated learning}\label{Federated}
Since of the privacy issue of EEG data, we leverage FL to train a model for accurate brain activity inference. By using FL, we can manage to train a global model without direct access to the raw training EEG data. 
Specifically, the techniques of FL follows a server-client setting. A server acts as model aggregator. In each round, the server collects updated feature mapping models and EEG classifiers from each client for model aggregation. 
Federated averaging \cite{DBLP:journals/corr/McMahanMRA16} is conducted over clients of each device for feature mapping aggregation and over all the clients for EEG classifier aggregation. 
After model aggregation, the server sends the updated global model to each client. When a client receives the model sent by server, it updates the model with its local data distributed on the projection manifold $\mathcal{P}_i (\mathcal{M}_i) \subset \mathcal{N} \hookrightarrow  \mathbb{R}^{F}$. The training process continues until the model converges.


\section{Experiments}
To verify the effectiveness of our approach, we conduct experiments on real-world EEG datasets. 
 
\subsection{MindBigData}
The \emph{MindBigData} dataset \footnote{The \emph{MindBigData} dataset is acquired and processed (\url{http://www.mindbigdata.com/opendb/index.html})} is a publicly available dataset containing millions of EEG brain signals of two seconds each, captured with the stimulus of seeing a digit from 0 to 9 and thinking about it, or captured without the stimulus of seeing the digits for contrast. Our task is to infer whether the subject receives a stimulus of seeing and thinking about a digit, or not, according to the corresponding EEG signal. There are three types of devices with varying channels and sampling frequencies adopted for the same task: 1) \emph{MindWave (MW)} with one channel and 512HZ, 2) \emph{EPOC (EP)} with 14 channels and 128HZ, and 3) \emph{Muse (MU)} with 4 channels and 220HZ. Federated learning over such heterogeneous dataset yields to a requirement of heterogeneous transfer learning for subspace alignment.

\subsection{Evaluation Metric and Methods}
The metric, \emph{accuracy}, is used to measure the effectiveness of our approach, and the following methods will be compared with our approach. 
\begin{itemize}
\item \emph{Baseline}: The neural network approach of classifier is trained on each device dataset. Three device dataset yield three baselines in total.
\item \emph{MU + MW} or \emph{MW + EP} or \emph{MU + EP}: Our approach, HHHFL method, is trained on every two of three device datasets. We call them directly the abbreviations of two datasets. 
\item \emph{MU + MW + EP}: Our approach, HHHFL method, is trained on three device datasets.
\end{itemize}

\subsection{Implementation Details and Result Analysis}
The experiments are conducted on a machine with 1.3 GHz Intel Core i5 with memory 4 GB 1600 MHz DDR3. We implement each private network architecture with convolutional neural network (CNN) layer and fully-connected layer (FC) with the dimensions of inputs 512, 440 and 1024 for three devices \emph{MU}, \emph{EP} and \emph{MW} respectively. The reduced dimension on the common embedding space $\mathcal{N}$ is 10. We adopt PySyft framework for the Horizontal Federated Learning (HFL). 

\begin{figure}[htp]
  \begin{minipage}[t]{0.3\textwidth}
  \center
  \includegraphics[width=1.7in]{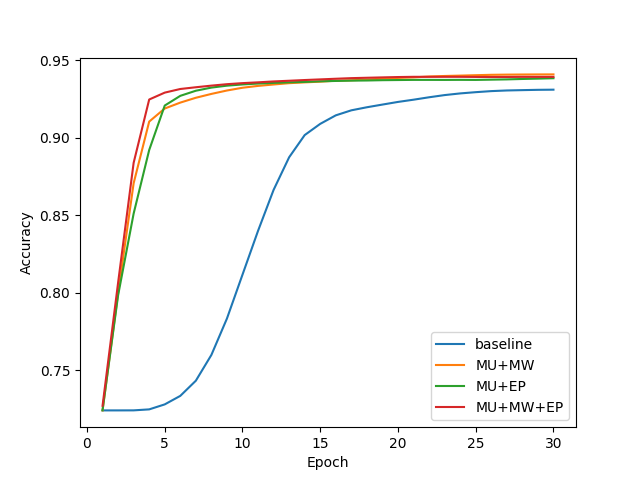} 
  \caption{HHHFL on MU}
  \label{Fig3}
\end{minipage}
\hspace{0.5cm}
\begin{minipage}[t]{0.3\textwidth}
 \center
  \includegraphics[width=1.7in]{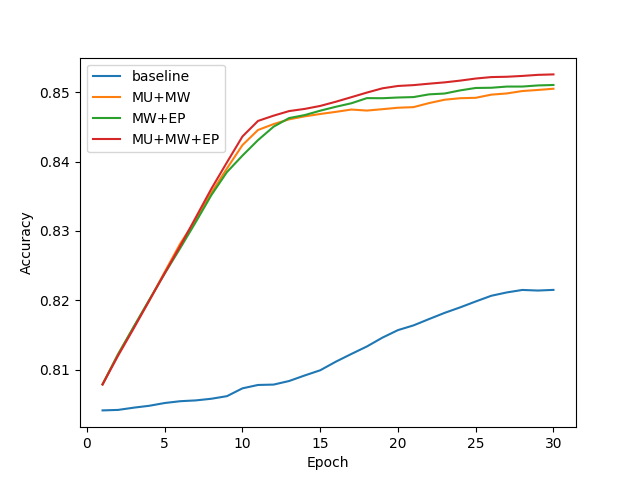} 
  \caption{HHHFL on MW}
  \label{Fig4}
  \end{minipage}
  \hspace{0.5cm}
\begin{minipage}[t]{0.3\textwidth}
  \center
  \includegraphics[width=1.7in]{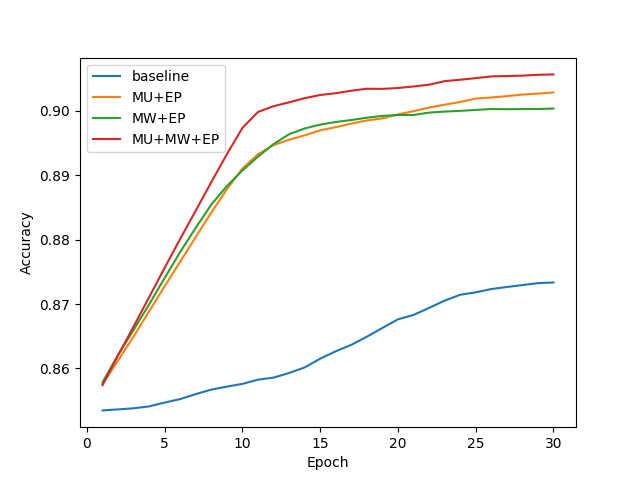} 
  \caption{HHHFL on EP}
  \label{Fig5}
 \end{minipage}
\end{figure}

Upon the performance on each figure above, we observe that our HHHFL approach, on two or three device datasets, outperform the baselines in accuracy and converge speed apparently. Adopting more data from the third device on our model yields a slight improvement in accuracy in Figure \ref{Fig4} and Figure \ref{Fig5}.

\section{Future Work}
We plan to further augment data privacy-preservation by applying secure multi-party computation and differential privacy. Secure multiparty computation can be applied to the proposed heterogeneous federated learning approach, to prevent information leakage during model training. Differential privacy will also be leveraged to prevent membership-inference attack from adversarial model querier.


\subsubsection*{Acknowledgments}
We would like to thank FedAI community for their helpful suggestions and contributions. 



{\footnotesize
\bibliographystyle{unsrt}
\bibliography{refs}

\begin{thebibliography}{10}

\bibitem{chong2013parameter}
Michelle Siu~Tze Chong.
\newblock {\em Parameter and state estimation of nonlinear systems with
  applications in neuroscience}.
\newblock PhD thesis, 2013.

\bibitem{subha2010eeg}
D~Puthankattil Subha, Paul~K Joseph, Rajendra Acharya, and Choo~Min Lim.
\newblock Eeg signal analysis: a survey.
\newblock {\em Journal of medical systems}, 34(2):195--212, 2010.

\bibitem{angelakis2007eeg}
Efthymios Angelakis, Stamatina Stathopoulou, Jennifer~L Frymiare, Deborah~L
  Green, Joel~F Lubar, and John Kounios.
\newblock Eeg neurofeedback: a brief overview and an example of peak alpha
  frequency training for cognitive enhancement in the elderly.
\newblock {\em The clinical neuropsychologist}, 21(1):110--129, 2007.

\bibitem{Li:2009:TAL:1631111.1631118}
Xiaowei Li, Bin Hu, Tingshao Zhu, Jingzhi Yan, and Fang Zheng.
\newblock Towards affective learning with an eeg feedback approach.
\newblock In {\em Proceedings of the First ACM International Workshop on
  Multimedia Technologies for Distance Learning}, MTDL '09, pages 33--38, New
  York, NY, USA, 2009. ACM.

\bibitem{yu2018building}
Han Yu, Zhiqi Shen, Chunyan Miao, Cyril Leung, Victor~R Lesser, and Qiang Yang.
\newblock Building ethics into artificial intelligence.
\newblock {\em arXiv preprint arXiv:1812.02953}, 2018.

\bibitem{regulation2016regulation}
General Data~Protection Regulation.
\newblock Regulation (eu) 2016/679 of the european parliament and of the
  council of 27 april 2016 on the protection of natural persons with regard to
  the processing of personal data and on the free movement of such data, and
  repealing directive 95/46.
\newblock {\em Official Journal of the European Union (OJ)}, 59(1-88):294,
  2016.

\bibitem{DBLP:journals/corr/McMahanMRA16}
H.~Brendan McMahan, Eider Moore, Daniel Ramage, and Blaise~Ag{\"{u}}era
  y~Arcas.
\newblock Federated learning of deep networks using model averaging.
\newblock {\em CoRR}, abs/1602.05629, 2016.

\bibitem{wang2011heterogeneous11}
Chang Wang and Sridhar Mahadevan.
\newblock Heterogeneous domain adaptation using manifold alignment.
\newblock In {\em Twenty-Second International Joint Conference on Artificial
  Intelligence}, 2011.

\bibitem{hwang2013eeg}
Han-Jeong Hwang, Soyoun Kim, Soobeom Choi, and Chang-Hwan Im.
\newblock Eeg-based brain-computer interfaces: a thorough literature survey.
\newblock {\em International Journal of Human-Computer Interaction},
  29(12):814--826, 2013.

\bibitem{cnnSchirr17}
R.~{Schirrmeister}, L.~{Gemein}, K.~{Eggensperger}, F.~{Hutter}, and T.~{Ball}.
\newblock Deep learning with convolutional neural networks for decoding and
  visualization of eeg pathology.
\newblock In {\em 2017 IEEE Signal Processing in Medicine and Biology Symposium
  (SPMB)}, pages 1--7, Dec 2017.

\bibitem{bird2019deep}
Jordan~J Bird, Diego~R Faria, Luis~J Manso, Anik{\'o} Ek{\'a}rt, and
  Christopher~D Buckingham.
\newblock A deep evolutionary approach to bioinspired classifier optimisation
  for brain-machine interaction.
\newblock {\em Complexity}, 2019, 2019.

\bibitem{6335751}
S.~{Gu} and Y.~{Jin}.
\newblock Heterogeneous classifier ensembles for eeg-based motor imaginary
  detection.
\newblock In {\em 2012 12th UK Workshop on Computational Intelligence (UKCI)},
  pages 1--8, Sep. 2012.

\bibitem{roesler2014comparison}
Oliver Roesler, Lucas Bader, Jan Forster, Yoshikatsu Hayashi, Stefan
  He{\ss}ler, and David Suendermann-Oeft.
\newblock Comparison of eeg devices for eye state classification.
\newblock {\em Proc. of the AIHLS}, 2014.

\bibitem{DBLP:journals/corr/KonecnyMRR16}
Jakub Konecn{\'{y}}, H.~Brendan McMahan, Daniel Ramage, and Peter
  Richt{\'{a}}rik.
\newblock Federated optimization: Distributed machine learning for on-device
  intelligence.
\newblock {\em CoRR}, abs/1610.02527, 2016.

\bibitem{liu2018secure}
Yang Liu, Tianjian Chen, and Qiang Yang.
\newblock Secure federated transfer learning, 2018.

\bibitem{NIPS2017_7029}
Virginia Smith, Chao-Kai Chiang, Maziar Sanjabi, and Ameet~S Talwalkar.
\newblock Federated multi-task learning.
\newblock In {\em NIPS}, pages 4424--4434. 2017.

\bibitem{wang2018heterogeneous18}
Xuesong Wang, Yuting Ma, Yuhu Cheng, Liang Zou, and Joel~JPC Rodrigues.
\newblock Heterogeneous domain adaptation network based on autoencoder.
\newblock {\em Journal of Parallel and Distributed Computing}, 117:281--291,
  2018.

\bibitem{duan2012learning}
Lixin Duan, Dong Xu, and Ivor Tsang.
\newblock Learning with augmented features for heterogeneous domain adaptation.
\newblock {\em arXiv preprint arXiv:1206.4660}, 2012.

\bibitem{gretton2007kernel}
Arthur Gretton, Karsten Borgwardt, Malte Rasch, Bernhard Sch{\"o}lkopf, and
  Alex~J Smola.
\newblock A kernel method for the two-sample-problem.
\newblock In {\em Advances in neural information processing systems}, pages
  513--520, 2007.

\end{thebibliography}
}

\newpage
\appendix
\section{Hierarchical Architecture}
\begin{figure}[htp]
  \center
  \includegraphics[width=0.75\textwidth]{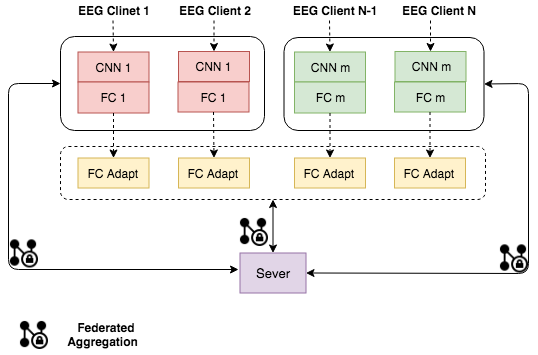} 
  \caption{The illustration demonstrates our proposed architecture HHHFL. Red and green block refer to two kinds of device data trained in typical horizontal FL. The yellow block is hierachical to the red and green block, which yields our architecture has the hierachical meaning.}
  \label{app}
\end{figure}

\end{document}